\documentclass[aps,10pt,twocolumn,preprintnumbers,prl,showpacs]{revtex4}

\usepackage{amsmath,amssymb,amsfonts,amsthm}
\usepackage{amsbsy} 
\usepackage{epsfig}
\usepackage{latexsym}
\usepackage{wasysym}

\usepackage{color}
\usepackage{hyperref}

\def\barr{\begin{array}}
\def\earr{\end{array}}

\def\ben{\begin{equation}}
\def\een{\end{equation}}
\def\bs{\begin{subequations}}
\def\es{\end{subequations}}
\def\bena{\begin{eqnarray}}
\def\eena{\end{eqnarray}}

\begin{document}

\title{Comment on ``Causality-violating Higgs singlets at the LHC''}
\author{Steffen Gielen}
\email{sgielen@perimeterinstitute.ca}
\affiliation{Riemann Center for Geometry and Physics, Leibniz Universit\"at Hannover, Appelstra\ss e 2, 30167 Hannover, Germany}
\affiliation{Perimeter Institute for Theoretical Physics, 31 Caroline St. N., Waterloo, Ontario N2L 2Y5, Canada} 

\begin{abstract}
The spacetime of Ho and Weiler [Phys. Rev. D {\bf 87}, 045004 (2013)] supposedly admitting closed timelike curves (CTCs) is flat Minkowski spacetime with a compactified coordinate and can only contain CTCs if the compact direction is chosen to be timelike. This case of a ``periodic time'' is probably the simplest example of a causality-violating spacetime; it trivially satisfies all energy conditions usually assumed in general relativity, and its geodesics are just straight lines. Its relevance for phenomenology of the LHC, on the other hand, depends on consistency with observational constraints on gravity, as is mentioned in general but not discussed in any detail by Ho and Weiler. We verify a basic consistency check for stationary sources. 
\end{abstract}

\preprint{pi-partphys-322}
\pacs{04.20.Cv, 04.20.Jb}

\maketitle

Solutions of general relativity that permit ``time travel'' on closed timelike curves (CTCs) have a long history, the G\"odel solution \cite{godel} being probably the most famous example. It is now established that CTCs occur in rather generic situations in general relativity, although the associated solutions often display singularities. Ho and Weiler \cite{ctcpaper} claim to construct a new class of such solutions in five spacetime dimensions, free of any pathologies, of the form
\bena
ds^2 &=& -d{\bf x}\cdot d{\bf x} + dt^2 + 2 g(u)\,dt\,du - h(u)\, du^2
\label{metric}
\\&=& -d{\bf x}\cdot d{\bf x} + (dt+g(u)\,du)^2 -(g^2(u)+h(u)) du^2\,,\nonumber
\eena
where $g^2+h>0$ everywhere so that the metric has signature $(+----)$; $u$ is compactified with period $L$. 

\begin{figure}[ht]
\begin{center}
\begin{picture}(90,120)
\put(20,10){\line(0,1){90}}\put(70,10){\line(0,1){90}}\put(45,0){$\vdots$}\put(45,100){$\vdots$}
\put(18,20){$\bullet$}\put(68,40){$\bullet$}
\put(18,40){$\circ$}\put(68,60){$\circ$}
\put(18,60){$\diamond$}\put(68,80){$\diamond$}
\put(58,20){\line(1,0){5}}\put(58,40){\line(1,0){5}}\put(60,25){\vector(0,-1){5}}\put(60,35){\vector(0,1){5}}\put(55,27){$L\overline{g}$}
\put(35,90){\vector(-1,0){15}}\put(55,90){\vector(1,0){15}}\put(38,87){$L\overline{D}$}
\put(0,10){\vector(1,0){10}}\put(0,10){\vector(0,1){10}}\put(10,10){$\overline{u}$}\put(0,20){$\overline{t}$}
\end{picture}
\end{center}
\caption{Flat spacetime with periodic identification.}
\label{fig}
\end{figure}
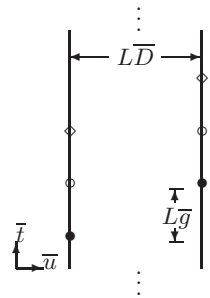

Does (\ref{metric}) admit CTCs? To answer this question it is helpful to change coordinates to $\overline{t}$ and $\overline{u}$ defined by
\ben
\overline{t} = t + \int\limits_0^{u}  du'\,g(u')\,,\quad\overline{u} = \int\limits_0^{u}  du'\,\sqrt{g^2(u')+h(u')}\,,
\een
so that (\ref{metric}) just becomes the flat Minkowski metric. In the new coordinates, the periodic identification of $u$ means that $(\overline{t},\overline{u})$ is to be identified with $(\overline{t}+L\,\overline{g},\overline{u}+L\,\overline{D})$, where
\ben
\overline{g} = \frac{1}{L}\int\limits_0^{L}  du\,g(u)\,,\quad\overline{D} = \frac{1}{L}\int\limits_0^{L}  du\,\sqrt{g^2(u)+h(u)}
\label{average}
\een
(in much of the analysis of \cite{ctcpaper} $\overline{D}=1$). The $(\overline{t},\overline{u})$ plane is depicted in Fig. \ref{fig}, where points specified by the same symbol are to be thought of as identified. 
Apart from these identifications, it is just a portion of flat spacetime, where geodesics are straight lines, and a curve can only be timelike if its tangent vector is always at an angle of less than $45^\circ$ with the vertical axis \footnote{Such a curve can also be null or spacelike since we are ignoring motion in the spacelike $x$ directions.}. It is then obvious that a CTC will exist if and only if $|\overline{g}|>\overline{D}$; in that case there will also be a closed timelike geodesic (so that the restriction to geodesics made in \cite{ctcpaper} is not relevant). By the same argument, closed null curves are possible already when $|\overline{g}|=\overline{D}$.

CTCs obviously exist in Minkowski spacetime with periodically identified timelike direction, but in \cite{ctcpaper} they seem to be more generic. Looking at Sec. IIIA in \cite{ctcpaper} one can see why: ``{\em By definition, a CTC is a geodesic that returns a particle to the same space coordinates from which it left, with an arrival time before it left.}'' But apart from the unnecessary restriction to geodesics, on a CTC one can return to the same point in space {\em and} time. 

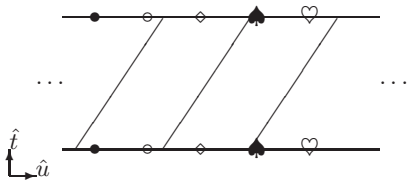
\begin{figure}[ht]
\begin{center}
\begin{picture}(120,80)
\put(10,20){\line(1,0){120}}\put(10,70){\line(1,0){120}}\put(0,45){$\hdots$}\put(130,45){$\hdots$}
\put(20,18){$\bullet$}\put(20,68){$\bullet$}
\put(40,18){$\circ$}\put(40,68){$\circ$}
\put(60,18){$\diamond$}\put(60,68){$\diamond$}
\put(80,18){$\spadesuit$}\put(80,68){$\spadesuit$}
\put(100,18){$\heartsuit$}\put(100,68){$\heartsuit$}
\put(15,20){\line(2,3){33}}\put(48,20){\line(2,3){33}}\put(81,20){\line(2,3){33}}
\put(-10,10){\vector(1,0){10}}\put(-10,10){\vector(0,1){10}}\put(0,10){$\hat{u}$}\put(-10,20){$\hat{t}$}
\end{picture}
\end{center}
\caption{Flat spacetime with a brane and causality violation.}
\label{fig2}
\end{figure}

The condition $|\overline{g}|>\overline{D}$ for the existence of CTCs can be derived from a modification of the analysis of Sec. III of \cite{ctcpaper}. Equation (3.18) of \cite{ctcpaper} states that on a geodesic, the coordinate $t$ takes the values
\ben
t_N\equiv t(u=\pm NL)=\pm\left(\frac{1}{\beta_0}-A\right)NL
\een
when crossing the brane for the $N$th time, starting from $t(0)=0$. For a closed curve we must clearly have $t_N=0$ for some $N$ and hence $\frac{1}{\beta_0}=A$. On the other hand, the geodesic can be timelike if (setting $g^2 + h = 1$ as in \cite{ctcpaper})
\ben
\left|g_0+\frac{1}{\beta_0}\right|>1\,.
\een
But $g_0+A=\overline{g}$ and hence we see that CTCs only exist if $|\overline{g}|>1$, or generally $|\overline{g}|>\overline{D}$, in which case one has compactified time in the 5D spacetime. This becomes more explicit after introducing coordinates in which the metric is the Minkowski metric, and only a single coordinate is compactified: After the coordinate transformation
\ben
\hat{t}=\frac{\overline{t}-\frac{\overline{D}}{\overline{g}}\overline{u}}{\sqrt{1-\frac{\overline{D}^2}{\overline{g}^2}}}\,,\quad \hat{u}=\frac{\overline{u}-\frac{\overline{D}}{\overline{g}}\overline{t}}{\sqrt{1-\frac{\overline{D}^2}{\overline{g}^2}}}\,,
\een
a boost in Minkowski spacetime which leaves the metric invariant, $(\hat{t},\hat{u})$ is to be identified with $\left(\hat{t}+L\,\overline{g}\,(1-(\overline{D}/\overline{g})^2)^{1/2},\hat{u}\right)$, so that the compactification is now restricted to $\hat{t}$. The brane supposedly corresponding to the Universe visible to us is then a ``diagonal'' timelike slice through this five-dimensional space, see Fig. \ref{fig2}. 

The 5D bulk spacetime with $|\overline{g}|>\overline{D}$ is certainly not a new solution to general relativity, but the most basic example of a causality-violating spacetime. The lengthy analysis of Secs. II--V of \cite{ctcpaper} which gives different forms of the metric, derives the geodesics, discusses the conditions for causality violation (missing the only relevant one, $|\overline{g}|>\overline{D}$) and shows that energy conditions are satisfied, is rather unnecessary since all these things are well known for Minkowski spacetime. The fact that CTCs only exist for $|\overline{g}|>\overline{D}$ must also be taken into account when reading the later calculations in \cite{ctcpaper}; in Sec. VII B the authors compute a dispersion relation for the Klein-Gordon field, assuming that $|\overline{g}|<\overline{D}$ in their calculation. Much of the phenomenological analysis in Sec. VII of \cite{ctcpaper}, despite appearance, only discusses spacetimes without causality violation.

The 4D brane does not contribute to the energy-momentum tensor, and hence has zero tension. It is then hardly surprising that the bulk spacetime is regular Minkowski spacetime and has no singularities.

What about the possible relevance of the 5D spacetime considered in \cite{ctcpaper} for phenomenology? The usual issues with causality-violating spacetimes are mentioned in \cite{ctcpaper}, but no time travel can occur for matter confined to the brane. However, as in other braneworld scenarios, gravity permeates all of the five spacetime dimensions and so one expects a noncompact fourth spatial dimension to be at odds with the Newtonian force law. In the ADD scenario, compatibility with observation constrains the size of extra dimensions, as is mentioned in Sec. VII A 2 of \cite{ctcpaper}; in the Randall-Sundrum model there is a noncompact dimension but a warped geometry which confines gravitons to the brane, as discussed in \cite{randsun}.

Perturbations of the 5D bulk spacetime (in transverse-traceless gauge) satisfy the wave equation in flat space,
\ben
\pentagon h_{AB}(x,\hat{t},\hat{u}) = -16\pi G\,T_{AB}(x,\hat{t},\hat{u})
\label{wave}
\een
where $T_{AB}$ is the 5D energy-momentum tensor of the perturbation, with periodic boundary conditions:
\ben
h_{AB}(x,\hat{t},\hat{u})=h_{AB}(x,\hat{t}+T,\hat{u})
\label{period}
\een
where $T:=L\,\overline{g}\,(1-(\overline{D}/\overline{g})^2)^{1/2}$ is the range of the periodic coordinate $\hat{t}$. Assuming, for instance, a point source stationary on the brane,
\ben
T_{AB}(x,\hat{t},\hat{u})\propto \sum_n\delta(x-x')\delta(\hat{u}-\gamma(\hat{t}-n\cdot T))
\een
where $\gamma$ specifies the ``tilt'' of the brane, and the ansatz 
\ben
h_{AB}(x,\hat{t},\hat{u})=\sum_n h_{AB}^{(n)}(x,\hat{u}-\gamma(\hat{t}-n\cdot T))\,,
\een
each $h_{AB}^{(n)}$ solves the 4D Poisson equation; the usual asymptotically vanishing solution falls off as
\ben
h_{AB}^{(n)}\propto\frac{1}{(x-x')^2+(\hat{u}-\gamma(\hat{t}-n\cdot T))^2}
\een
so that $h_{AB}=\sum_n h_{AB}^{(n)}$ solves (\ref{wave}) and (\ref{period}). Taking $\hat{u}$ to be also on the brane so that $\hat{u}=\gamma(\hat{t}+ n_0 T)$ for some $n_0$, the sum over $n$ can be computed exactly and one obtains
\ben
h_{AB}\propto\frac{\pi\coth(\pi(x-x')/\gamma T)}{(x-x')\gamma T}\,;
\een
for $|x-x'|\gg\gamma T$ the hyperbolic cotangent is essentially one and a $1/r$ potential emerges, as required. 

For a source local in space and time, this argument does not apply and the resulting field strongly deviates from what is seen when only three dimensions are noncompact, being given by a Green's function of the 5D, not the 4D flat space wave equation. Clearly, more work is needed to check the consistency of the setup in \cite{ctcpaper} for particle phenomenology, or even LHC.

{\em Acknowledgments.} I thank T. J. Weiler and the referee for helpful comments which clarified the picture. I was supported by a Riemann Fellowship of the Riemann Center for Geometry and Physics. Research at Perimeter Institute is supported by the Government of Canada through Industry Canada and by the Province of Ontario through the Ministry of Research and Innovation.

\end{document}